\documentclass
[superscriptaddress,secnumarabic,amssymb,amsmath,nobibnotes,aps,prd,showkeys,showpacs,twocolumn,nofootinbib]{revtex4-1}%

\usepackage{bm}
\usepackage{mathrsfs}
\usepackage[all]{xy}
\usepackage{epsfig,subfigure}
\usepackage{latexsym, amssymb, amsmath, amsfonts}
\usepackage[english]{babel}
\usepackage[OT1]{fontenc}
\usepackage[utf8]{inputenc}
\usepackage{makeidx}
\usepackage[colorlinks=true,breaklinks=true]{hyperref}
\usepackage{graphicx, wrapfig, epsf}
\usepackage{float}
\usepackage[dvipsnames, table]{xcolor}
\hypersetup{urlcolor=BlueViolet,
	        citecolor=Plum,
	        linkcolor=OliveGreen}

\definecolor{Gray}{rgb}{.9,.9,.9}

\begin{document}

\title{Jupiter and jovian (exo)-planets in Palatini $f(\bar R)$ gravity}

\author{Aneta Wojnar}
\email{aneta.magdalena.wojnar@ut.ee}
\affiliation{Laboratory of Theoretical Physics, Institute of Physics, University of Tartu,
W. Ostwaldi 1, 50411 Tartu, Estonia
}

\begin{abstract}
Some parts of the substellar evolution, such as fragmentation of a gaseous cloud and Jupiter-like planet's cooling, are demonstrated to be impacted by Palatini $f(\bar R)$ gravity. Using simple models describing those processes we show that the opacity mass limit as well as cooling time of jovian planets differ in modified gravity.
\end{abstract}

\maketitle

\section{Introduction}
Our understanding of the surrounding world was significantly enriched by the Einstein's proposal \cite{ein1,ein2}, the theory of General Relativity (GR), which has been subsequently tested by many observations and experiments \cite{Will:2014kxa}. There is no doubt that the most spectacular one among them is the confirmation of the black holes' existence by the gravitational waves' detection coming from a merger of two such objects \cite{TheLIGOScientific:2017qsa}, and by the direct observation of the black hole's shadow from the center of the M87 galaxy \cite{Akiyama:2019cqa,aki2,aki3,god}
(see \cite{Barack:2018yly} for review). However, in order to explain some of the cosmological and astrophysical phenomena for which GR does not provide satisfactory explanations, many other gravitational models have been proposed to shed light on the nature of dark matter and dark energy  \cite{Copeland:2006wr,Nojiri:2006ri,nojiri2,nojiri3,Capozziello:2007ec,Carroll:2004de}, spacetime singularities \cite{Senovilla:2014gza}, unification of physics of different scales \cite{ParTom,BirDav}, as well as the existence of massive compact objects exceeding theoretical predictions  \cite{lina,as,craw,NSBH,abotHBH,sak3}.

One of the features of some of those proposals is the fact that they modify the non-relativistic limit of equations describing stellar and substellar objects, by for example introducing terms which in particular cases can be expressed by the functions of energy density \cite{Saito:2015fza,olek2,olmo_ricci} (for a review see \cite{review,cantata}). Such a property has provided different limiting masses for various kinds of astrophysical objects, among which one can distinguish the Chandrasekhar mass for white dwarf stars \cite{Chandra,Saltas:2018mxc,Jain:2015edg,Banerjee:2017uwz,Wojnar:2020wd,Belfaqih:2021jvu}, the minimum Main Sequence mass \cite{sak1,sak2,Crisostomi:2019yfo,gonzalo}, or minimum mass for deuterium burning \cite{rosyadi}. It turn out that those theories also alter the stellar early and post-Main Sequence evolution \cite{aneta2, chow}; they also have impact on cooling processes of brown dwarfs \cite{maria} and alter lithium abundances in stellar atmospheres \cite{aneta3}.

Moreover, it was also demonstrated that gravitational theories different than GR can have a non-negligible impact on terrestrial (exo-)planets' profiles, providing a possibility to test such theories with the use of seismic data \cite{olek}. Since such planets are much smaller and their gravitational fields are weaker \footnote{And because of that fact the effects of modified gravity can be hidden in the observational uncertainties.} than those of gaseous giants from our or other planetary systems, let us turn our attention to the jovian planets. The gaseous giant planets possess, regarding their inner structure, many similarities to the bigger substellar objects, mainly brown dwarf stars. Acknowledging that the modified gravity impact on brown dwarfs' properties could be detectable by our current technology \cite{sak1,sak2,Crisostomi:2019yfo,gonzalo,rosyadi,maria}, Jupiter and Jupiter-like (exo-)planets may also provide an excellent opportunity to understand the gravity effects on dense environments. 

Before doing so, let us recall the basic notions regarding Palatini $f(\bar R)$ gravity, which will be our modified gravity framework. We will use it to demonstrate that the evolution of jovian planets can slightly differ than in the one provided by GR. In contrary to the metric approach, an independent connection is introduced, which arises to the fact that we deal with two independent geometric structures: metric $g$, and the connection $\Gamma$. It was demonstrated in various works that this approach carries a number of advantages \cite{junior,sch,ol1,ol2} which we will not discuss here, though. 
Let us also comment that in this work we use $(-+++)$ signature convention while $\kappa=-8\pi G/c^4$.

The action of $f(\bar{R})$ gravity in Palatini formulation is given by
\begin{equation}
S[g,\Gamma,\psi_m]=\frac{1}{2\kappa}\int \sqrt{-g}f(\bar{R}) d^4 x+S_{\text{matter}}[g,\psi_m],\label{action}
\end{equation}
where $\bar{R} = g^{\mu\nu}\bar{R}_{\mu\nu}(\Gamma)$ is the Palatini curvature scalar, built from the metric and the independent connection, while $\psi_m$ denotes matter fields. The variation of the action is taken with respect to both structures; the metric one gives 
\begin{equation}
f'(\bar{R})\bar{R}_{\mu\nu}-\frac{1}{2}f(\bar{R})g_{\mu\nu}=\kappa T_{\mu\nu},\label{structural},
\end{equation}
where $T_{\mu\nu}=-\frac{2}{\sqrt{-g}}\frac{\delta S_m}{\delta g_{\mu\nu}}$ is the energy-momentum tensor, and prime is understood here as differentiating with respect to the curvature. Contracting the above equation with the metric $g_{\mu\nu}$ provides an algebraic relation between the Palatini curvature and the trace of energy-momentum tensor:
\begin{equation}
    f'(\bar{R})\bar{R}-2f(\bar{R})=\kappa T.
\end{equation}
This feature allows to solve the above equations in some particular choices of the functional $f(\bar{R})$, providing that $\bar R=\bar R(T)$.

On the other hand, the relation between the connection and the metric tensor is given by the variation of (\ref{structural}) with respect to $\Gamma$, which can be written as
\begin{equation}
\nabla_\beta(\sqrt{-g}f'(\bar{R}(T))g^{\mu\nu})=0.\label{con}
\end{equation}
The above covariant derivative is understood as the one defined by the independent connection. Defining a metric tensor $h_{\mu\nu}$ such that
\begin{equation}
    h_{\mu\nu} = f'(\bar{R}(T))g_{\mu\nu}
\end{equation}
is conformally related to the metric $g_{\mu\nu}$, the equation \eqref{con} can be expressed as
\begin{equation}
\nabla_\beta(\sqrt{-h}h^{\mu\nu})=0.\label{con2}
\end{equation}
The connection $\Gamma$ happens to be the Levi-Civita one with respect to the metric $h_{\mu\nu}$ and to be an auxiliary field that can be integrated out, resulting that the degrees of freedom are related to the metric tensor $g_{\mu\nu}$ \cite{DeFelice:2010aj,BSS,SSB}.

The further parts of the paper are as follows: in the next section we will provide the basic equations describing spherical-symmetric objects in Palatini $f(\bar{R})$ gravity in the non-relativistic limit. Then, we will review the Jeans criterion for the given gravity model and we will show that metric-affine gravity also affects the fragmentation process, which is followed by the altered opacity mass limit, often used as a boundary value between brown dwarf stars and giant planets. In the second part of this work we will study jovian planet's evolution which turns out to differ in modified gravity framework, too. In the last section we will draw brief conclusions.

\section{Stellar and substellar toolkit}\label{toolkit}
In this section we will recall for the reader's convenience the basic equations which are needed to study low-mass stars and substellar objects, such as brown dwarfs and giant (exo-)planets. For the basic literature, see e.g. \cite{stellar,hansen,planets,planets2}; in what follows, we will focus on equations for the quadratic Palatini $f(\bar R)$ gravity derived in \cite{aneta1,artur,gonzalo,aneta2,aneta3,maria,olek}. Let us comment that in other theories of gravity some of the equations can be modified in a different way, or even they may be the same as in Newtonian physics.

For the Starobinski model 
\begin{equation}
 f(\bar{R})=\bar{R}+\beta\bar{R}^2,
\end{equation}
in the framework of the Palatini $f(\bar R)$ gravity, a spherical-symmetric low-mass object\footnote{We consider a toy-model of a star or planet, therefore we do not take into account on this stage of the work nonsphericity, magnetic fields, time-dependency,..., which require a numerical approach in order to consider more realistic stellar and substellar objects.} 
 is described by the hydrostatic equilibrium equations 
\begin{align}
 \frac{dp}{d\tilde r}&=-\frac{G m(\tilde r)\rho(\tilde r)}{\Phi(\tilde r)\tilde r^2} \ ,\\
 m&=\int_0^{\tilde r}4\pi x^2\rho(x)dx \ ,
\end{align}
where $\tilde r^2=\Phi(\tilde r) r^2$ and $\Phi(\tilde r)\equiv f'(\bar{R}(T))=1+2\kappa c^2\beta \rho(\tilde r)$. The radius coordinate $\tilde r$ is the one of the Einstein frame \cite{aneta1,aneta4}; transforming back to the 
physical (Jordan) frame and considering only the terms linear in $\kappa c^2\beta$, the modified hydrostatic equilibrium equation is
\begin{equation}\label{pres}
 p'=-g\rho(1+\kappa c^2 \beta [r\rho'-3\rho]) \ ,
\end{equation}
where prime denotes now the derivative with respect to the radius coordinate $r$, while $g$ is the surface gravity. In the further part we will approximate it on the planet's atmosphere as a constant since we may assume that $r_\text{atmosphere}\approx R$, where $R$ is the radius of the planet:
\begin{equation}\label{surf}
 g\equiv\frac{G m(r)}{r^2}\sim\frac{GM}{R^2}=\text{constant},
\end{equation}
where $M=m(R)$. Although the transformation of the mass function $ m(\tilde r)$ to $m(r)$ depends on the energy density which on the planet's surface will drop to zero, we are using in this work the well-known expression\footnote{See \cite{olek2,olek} for the modified one in that model.}
\begin{equation}
    m'(r)=4\pi r^2\rho(r).
\end{equation}
Using $m''=8\pi r\rho+4\pi r^2 \rho'$ and (\ref{surf}), we may rewrite (\ref{pres}) as
\begin{equation}\label{hyd}
 p'=-g\rho\left( 1+8\beta\frac{g}{c^2 r} \right).
\end{equation}

Another crucial element for the star's or planet's modelling is the heat transport in their interiors and atmospheres. The most common criterion which decides what kind of energy transport takes place is the Schwarzschild one \cite{schw,schw2}:
\begin{align*}
 \nabla_\text{rad}\leq&\nabla_\text{ad}\;\;\text{\small pure diffusive radiative or conductive transport}\\
  \nabla_\text{rad}>&\nabla_\text{ad}\;\;\text{\small adiabatic convection is present locally}
\end{align*}
where the gradient describing the temperature $T$ variation with depth is defined as follows
\begin{equation}
 \nabla_{\text{rad}}:=\left(\frac{d \ln{T}}{d\ln{p}}\right)_{\text{rad}}.
\end{equation}
It was demonstrated that in the Palatini case the Schwarzschild criterion is modified, since the temperature gradient is \cite{aneta2}
\begin{equation}\label{grad}
  \nabla_{\text{rad}}=\frac{3\kappa_{rc}lp}{16\pi acG mT^4}\left(1+8\beta\frac{G m}{c^2 r^3}\right)^{-1},
\end{equation}
where $l$ is the local luminosity, the radiation density constant is $a=7.57\times 10^{-15}\frac{erg}{cm^3K^4}$ and the opacity 
$\kappa_{rc}^{-1}=\frac{1}{\kappa_{rad}}+\frac{1}{\kappa_{cd}}$ 
with $\kappa_{rad}$ being the radiative opacity while $\kappa_{cd}$ is the conductive one.
The modification, depending on the sign of the parameter $\beta$, has a stabilizing or destabilizing effect. Putting $\beta=0$ recovers the standard Schwarzschild criterion. On the other hand, the adiabatic gradient $\nabla_\text{ad}$, as discussed further, is a constant value for particular cases.

A good approximation for the microscopic description of matter is given by the simple power-law relation between pressure and density, called polytropic equation of state (EoS)
\begin{equation}\label{pol}
  p=K\rho^{1+\frac{1}{n}},
\end{equation}
where $K$ depends on the composition of the fluid and may also carry information about the interactions between particles, the effects of electron degeneracy, and phase transitions, just to
mention a few phenomena that can be taken into account \cite{aud}, while $n$ is the polytropic index whose value describes different objects \cite{politropia}. In the following part of the paper, we will use the simplest relation in the case of fully convective objects, that is, their interior can be modelled by non-relativistic degenerate
electron gas\footnote{Examples of fully convective objects are low-mass stars with masses $\lesssim0.6M_\odot$, brown dwarfs, and giant gaseous planets; however, when their atmospheres are considered, one deals with radiative heat transport instead.}. Therefore, for the polytropic index $n=3/2$ one deals with the constant value (see e.g. \cite{stellar} for more details)
\begin{equation*}
    K=\frac{1}{20}\left(\frac{3}{\pi}\right)^\frac{2}{3}\frac{h^2}{m_e}\frac{1}{(\mu_e m_u)^\frac{5}{3}}.
\end{equation*}

In the case of the polytropes one uses a suitable approach, called the Lane-Emden formalism, allowing to rewrite the all relevant equations in the dimensionless form. For our particular model of gravity, the equation \eqref{hyd} is now transformed into the modified Lane-Emden equation \cite{aneta1}
\begin{equation}\label{LE}
 \frac{1}{\xi}\frac{d^2}{d\xi^2}\left[\sqrt{\Phi}\xi\left(\theta-\frac{2\alpha}{n+1}\theta^{n+1}\right)\right]=
 -\frac{(\Phi+\frac{1}{2}\xi\frac{d\Phi}{d\xi})^2}{\sqrt{\Phi}}\theta^n,
\end{equation}
where $\Phi=1+2\alpha \theta^n$ with $\alpha=\kappa c^2\beta\rho_c$, while the dimensionless $\theta$ and $\xi$ are defined as
\begin{align}
 r&=r_c\bar{\xi},\;\;\;\rho=\rho_c\theta^n,\;\;\;p=p_c\theta^{n+1},\label{def1}\\
 r^2_c&=\frac{(n+1)p_c}{4\pi G\rho^2_c},\label{def2}
\end{align}
where $p_c$ and $\rho_c$ are the core values of pressure and density, respectively.
The (numerical) solutions of the Lane-Emden equation (\ref{LE}) can be used to express star's mass, radius, central density, and temperature as
\begin{align}
 M&=4\pi r_c^3\rho_c\omega_n,\label{masss}\\
 R&=\gamma_n\left(\frac{K}{G}\right)^\frac{n}{3-n}M^\frac{1-n}{n-3} \label{radiuss},\\
 \rho_c&=\delta_n\left(\frac{3M}{4\pi R^3}\right) \label{rho0s} ,\\
 T&=\frac{K\mu}{k_B}\rho_c^\frac{1}{n}\theta_n \label{temps},
\end{align}
where $k_B$ is Boltzmann's constant and $\mu$ the mean molecular weight. It should be noticed that the constants (\ref{omega}) and (\ref{delta}) appearing in the above equations 
\begin{align}
 \omega_n&=-\frac{\xi^2\Phi^\frac{3}{2}}{1+\frac{1}{2}\xi\frac{\Phi_\xi}{\Phi}}\frac{d\theta}{d\xi}\mid_{\xi=\xi_R},\label{omega}\\
  \gamma_n&=(4\pi)^\frac{1}{n-3}(n+1)^\frac{n}{3-n}\omega_n^\frac{n-1}{3-n}\xi_R,\label{gamma}\\
 \delta_n&=-\frac{\xi_R}{3\frac{\Phi^{-\frac{1}{2}}}{1+\frac{1}{2}\xi\frac{\Phi_\xi}{\Phi}}\frac{d\theta}{d\xi}\mid_{\xi=\xi_R}} \ . \label{delta}
\end{align}
 include extra terms with respect to their well-known forms in GR/Newtonian physics \cite{artur}. 
 
 Using the defined quantities in the equations~\eqref{hyd} and (\ref{grad}) we may write them as hydrostatic equilibrium equation for polytropes
 \begin{equation}\label{hyd_pol}
 p'=-g\rho\left( 1+\frac{4\alpha}{3\delta} \right),
\end{equation}
while the Schwarzschild criterion is
\begin{equation}
 \nabla_{\text{rad}}=\frac{3\kappa_{rc}lp}{16\pi acG mT^4}\left(1-\frac{4\alpha}{3\delta_n}\right)^{-1}.
\end{equation}

If one deals with an object massive enough to burn light elements in its core, such as for instance hydrogen, deuterium, and lithium, the luminosity produced by this energy generation process is given by
\begin{equation}
    \frac{dL_\text{burning}}{dr}=4\pi r^2\dot\epsilon\rho,
\end{equation}
where the energy generation rate $\dot\epsilon$ is a function of energy density, temperature, and stellar composition, and it is often approximated as a power-low function of the two first \cite{fowler}. In our current work we will not study energy generation in the object's core, which for an equilibrium configuration is compensated by energy radiated from the surface. To see the works considering such processes in metric-affine gravity, see \cite{gonzalo,aneta3,rosyadi}.

The energy radiated  through the surface is given by the Stefan-Boltzmann law 
\begin{equation}\label{stefan}
    L=4\pi f\sigma T_\text{eff}^4R^2,
\end{equation}
where $\sigma$ is the Stefan-Boltzmann constant while $f$ is a factor with the value less than one in order to take into account that the object can radiate less than a black-body with the same effective temperature $T_\text{eff}$. To determine the effective temperature as well as to find some particular quantities in the atmosphere, one often uses the definition of the optical depth $\tau$  with a mean opacity $\bar\kappa$ (averaged over the stellar or planetary atmosphere, see e.g. \cite{stellar,hansen}):
\begin{equation} \label{eq:od}
 \tau(r)=\bar\kappa\int_r^\infty \rho dr.
\end{equation}
In further part of the work, as we will deal with low temperatures in atmospheres, we will use Rosseland mean opacity given by the Kramer law
\begin{equation}\label{abs}
 \bar\kappa= \kappa_0 p^u T^{4w},
\end{equation}
where $\kappa_0$, $u$ and $w$ are values depending on different opacity regimes \cite{kley,armitage}.

\section{The Jeans and opacity mass limits in Palatini $f(\bar R)$ gravity}

The Jeans mass is a critical mass of a gaseous cloud or of its fragment, which is still stable against gravitational collapse \cite{jeans}. Exceeding this mass, the cloud contracts until there appears some other process producing pressure balancing the gravitational one (such as for example electron degeneracy or pressure related to the hydrogen ignition) which stops the collapse.

It was demonstrated that the Jeans mass differs in some theories of gravity as for example metric $f(R)$ gravity \cite{capojeans}: for spherical-symmetric large clouds of gas modified gravity can be repulsive, causing that instead of forming quasi uniform bodies thin shells are produced \cite{arbuzowa}, and also that one deals with a faster growth of perturbations \cite{arbuzowa2}. Those results were used as well to constrain the $f(R)$ gravity model with Bok globules data \cite{vainio}.

It turns out that also dark matter models affect this mass limit, having various effects on structure formations in different astrophysical and intergalactic scales \cite{roshan,kremer,kremer2}. Quantum effects, such as e.g. extended or generalised uncertainty principle increase or decrease, respectively, the Jeans mass \cite{mora}. Other models, such as energy–momentum-squared \cite{kazemi} and non-minimal matter-curvature coupling gravities \cite{gomes} can also lead to changes in the Jeans criterion.

Jeans mass limit was also studied in metric-affine models of gravity, for instance Eddington inspired Born-Infeld gravity \cite{yang, martino, roshan2}, proving a departure from the standard scenario of self-gravitating systems, such as collisionless clouds or thin disks, or Palatini $f(\bar R)$ gravity \cite{raila}. In what follows, we will perform a simplified procedure which is better suited for our further purposes.

\subsection{The Jeans criterion}

To derive the Jeans criterion for the considered gravity model, we need to be equipped with Poisson \cite{alejandro}, Euler \cite{junior}, and continuity equations modified by the quadratic Palatini $f(\bar R)$ gravity:

\begin{align}
    \nabla^2\phi&=\frac{\kappa}{2}(\rho +2\beta\nabla^2\rho),\label{poisson}\\
    \rho(\partial_t\mathbf{v}+v^i\nabla_iv^j)&=-\rho\partial^j\phi-\partial^j\left(p+\frac{\kappa\beta\rho^2}{2}\right),\label{euler}\\
    0&=\partial_t\rho +\partial_j(\rho\, v^j)\label{continuity}
\end{align}
Therefore, let us consider a simplified situation when we are dealing with an infinite, homogeneous gas which obeys the above equations. 
For the equilibrium state we assume that 
\begin{equation}
    \rho=\rho_0=\text{const},\;\;T=T_0=\text{const},\;\;\mathbf{v}_0=\mathbf{0}=\text{const},
\end{equation}
while $\phi_0$ can be obtained by $\nabla^2\phi_0=4\pi G\rho_0$ and boundary conditions at infinity.

Moreover, the gas can be described by the ideal gas equation of state in the terms of isothermal speed sound, which we denote by $v_s$,
\begin{equation}
    p=\frac{\mathcal{R}}{\mu}\rho T = v^2_s \rho.
\end{equation}
We will perturb the above EoS together with the equations (\ref{poisson})-(\ref{continuity}) by the standard procedure
\begin{align}
    \rho=\rho_0+\rho_1,\;\;p=p_0+p_1,\;\;\phi=\phi_0+\phi_1,\;\;\mathbf{v}=\mathbf{v}_1,
\end{align}
where the quantities with the index $1$ depend on time and spatial coordinates. Up to the linear terms, the equations (\ref{poisson})-(\ref{continuity}) take the following forms\footnote{We have assumed that the perturbations are isothermal, therefore $v_s$ is not perturbed.}
\begin{align}
     \nabla^2\phi_1&=\frac{\kappa}{2}(\rho_1 +2\beta\nabla^2\rho_1),\label{poisson1}\\
    \partial_t\mathbf{v}_1=&-\nabla\left(\phi_1+v_s^2\frac{\rho_1}{\rho_0}+\frac{\kappa\beta}{2}\rho_1\right),\label{euler1}\\
    0&=\partial_t\rho_1 +\rho_0\partial_j v^j,\label{continuity1}
\end{align}
where the equilibrium terms denoted by the index $0$ vanished. Proceeding as usual, that is, assuming that for the above linear homogeneous system of differential equations there exists a solution of the form $\sim\text{exp}[i(kx+\omega t)]$ we end up with the relation 
\begin{equation}
    \omega^2=k^2\left(v_s^2+\kappa c^2\frac{\beta}{2}\rho_0\right)+\frac{\kappa}{2}\rho_0.
\end{equation}
We immediately notice that in Palatini $f(\bar R)$ gravity for $k\rightarrow\infty$ we deal with a reduced/higher isothermal sound waves\footnote{Let us recall that here and everywhere in the paper we use the negative convention for the constant $\kappa$.}. Therefore, it follows that the characteristic wave number $k_J$ is obtained by setting $\omega=0$ 
\begin{equation}
    k_J^2=-\frac{\kappa\rho_0}{2\left(v^2_s+\kappa c^2\frac{\beta}{2}\rho_0\right)}
\end{equation}
The perturbations are unstable when $k<k_J$ and stable otherwise. Defining a characteristic wavelenght by $\lambda_J:=2\pi/k_J$, we may write down the Jeans criterion for instability in quadratic Palatini gravity which reads
\begin{equation}
    \lambda>\lambda_J=\left(\frac{\pi\left(v^2_s+\kappa c^2\frac{\beta}{2}\rho_0\right)}{G\rho_0}\right)^\frac{1}{2}.
\end{equation}
Depending on the sign of the parameter $\beta$, Palatini gravity introduces stabilizing or destabilizing effect to that criterium.

\subsection{Virial theorem and Jeans mass}
Let us assume that the considered isothermal sphere of the ideal gas is embedded in the medium of a non-zero pressure. The virial theorem for non-vanishing surface pressure $p_0$ for the sphere is given by
\begin{equation}
    \int^M_0\frac{Gm}{r}\left(1-\frac{4\alpha}{3\delta}\right)dm=3\int^M_0\frac{p}{\rho}dm-4\pi R^3p_0.
\end{equation}
Using the Lane-Emden equation (\ref{LE}) and a perfect monatomic gas to integrate the above we may write 
\begin{equation}
    \chi E_i + E_g = 4\pi R^3 p_0,
\end{equation}
where for the ideal monotomic gas $\chi=2$, $E_i=c_v MT$ is the inertial energy and $E_g=-\frac{3}{5-n}\frac{GM^2}{R}\left(1-\frac{4\alpha}{3\delta}\right)$. Therefore, the surface pressure $p_0$ is given by
\begin{equation}
    p_0=\frac{c_vMT}{2\pi R^3}-\frac{\Theta GM^2}{4\pi R^4}\left(1-\frac{4\alpha}{3\delta}\right),
\end{equation}
where we have set $\Theta=3/(5-n)$. Introducing two scaling factors $\tilde R=\Theta GM/(2c_vT)$ and $\tilde p=c_vMT/(2\pi \tilde R^3)$ such that
\begin{equation*}
    R=x\tilde R,\;\;\; p_0=y\tilde p_0
\end{equation*}
one writes
\begin{equation}\label{igrek}
    y=\frac{1}{x^3}\left( 1-\frac{1-\frac{4\alpha}{3\delta}}{x} \right).
\end{equation}

\begin{figure}[t]
\centering
\includegraphics[scale=.65]{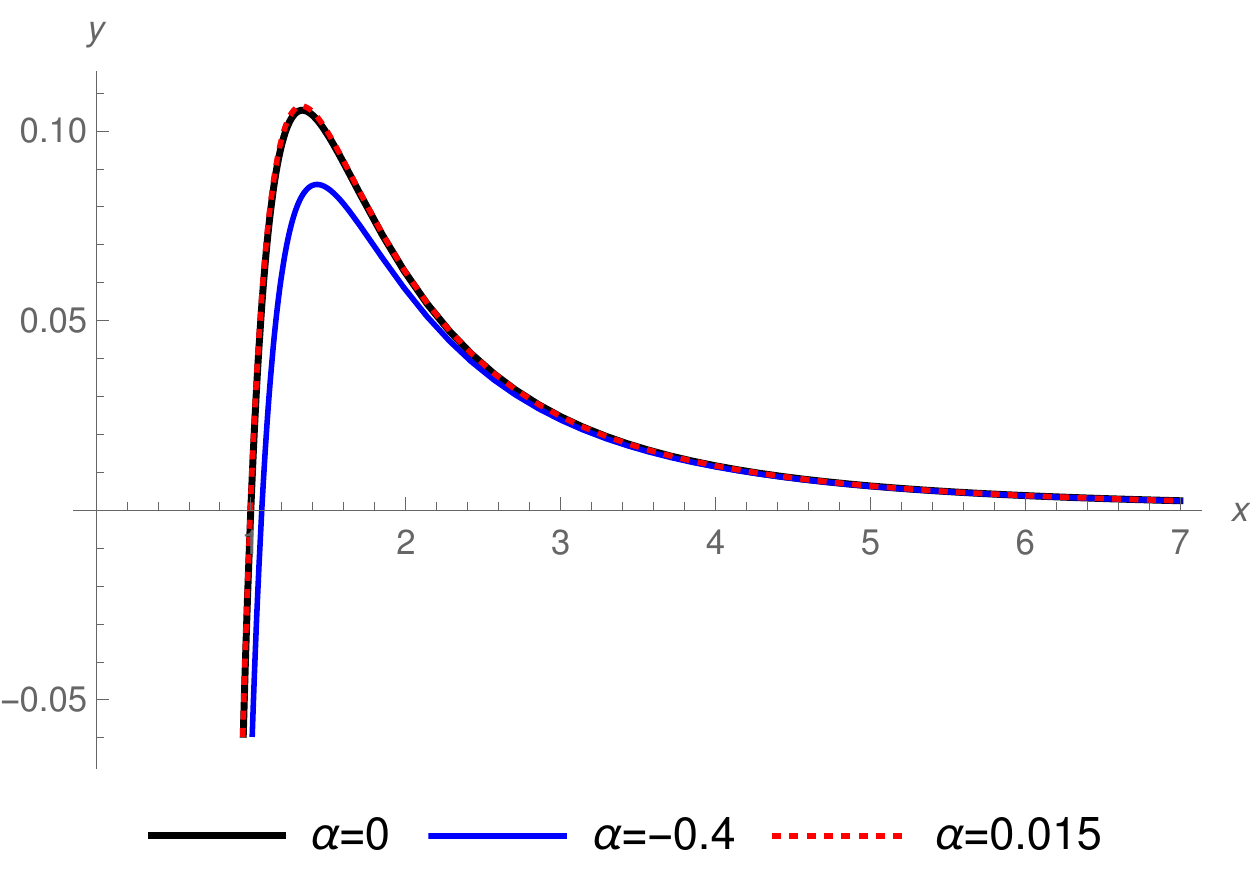}
\caption{[color online] The function (\ref{igrek}) representing the behaviour of dimensionless pressure $y$ with respect to the dimensionless radius $x$ for three different values of the parameter $\alpha$, which are in agreement with the constraints given in \cite{gonzalo}. $\alpha=0$ corresponds to the GR/Newtonian case. }
\label{fig.1}
\end{figure}

This function is depicted in the figure \ref{fig.1} for a few values of $\alpha$. One immediately notices how the function $y$ (pressure) changes from negative values to the positive ones with increasing dimensionless radius; after reaching its maximum the pressure approaches zero. This behaviour is however slightly different in modified gravity since the gravitational energy differs with respect to the GR/Newtonian case. It can be shown that the maximum of the pressure occurs for the radius
\begin{equation}\label{critR}
    R_m=\frac{4\Theta}{9}\frac{G\mu M}{\mathcal{R}T}\left(1-\frac{4\alpha}{3\delta}\right).
\end{equation}
Therefore, we deal with a stable configuration when $R>R_m$ while the Jeans instability is recovered for $R<R_m$ when $\alpha=0$. To see it, let us replace the mass by
\begin{equation}\label{mass}
M=4\pi R^3_m\bar\rho/3,
\end{equation}
where $\bar\rho$ is the mean density of the sphere. Then, $R_m$ is the critical radius of a gaseous mass of mean density $\bar\rho$ and temperature $T$ which is marginally stable:
\begin{equation}\label{critR2}
    R_m^2=\frac{27}{16\pi\Theta}\frac{\mathcal{R}T}{G\mu\bar\rho\left(1-\frac{4\alpha}{3\delta}\right)}
\end{equation}
which is of the same order (when $\alpha=0$) as the critical Jeans wavelength
\begin{equation}
\lambda_J^2=\frac{\pi\left(\frac{\mathcal{R}T}{\mu}+\kappa c^2\frac{\beta}{2}\bar\rho\right)}{G\rho_0},
\end{equation}
where we have used $v^2_s=\frac{\mathcal{R}T}{\mu}$.

In other words, each equilibrium state with the surface pressure $p_0$ and radius $R$ has its critical mass $M_J$. Since $R_m$ grows linearly with $M$ (\ref{critR}), masses larger than $M_J$ are not gravitationally stable, so when compressed a bit more, they will fall together. Using (\ref{critR2}) in $ M_J=\frac{4\pi}{3}\bar\rho R^3_m$ one obtains the so-called Jeans mass
\begin{equation}
    M_J=
    \frac{27}{16}\left(\frac{3}{\pi}\right)^\frac{1}{2}\left(\frac{\mathcal{R}}{\Theta G}\right)^\frac{3}{2}\left(\frac{T}{\mu}\right)^\frac{3}{2}\left(\frac{1}{\bar\rho}\right)^\frac{1}{2}\left(1-\frac{4\alpha}{3\delta}\right)^{-\frac{3}{2}}.
\end{equation}
Using the constants' values and rescaling the most crucial ingredients in the above expression, the Jeans mass can be rewritten as:
\begin{equation}\label{jeansmass}
    M_J=\frac{1.1 M_\odot}{\left(1-\frac{4\alpha}{3\delta}\right)^{\frac{3}{2}}}\left(\frac{T}{10\text{K}}\right)^\frac{3}{2}\left(\frac{\rho}{10^{-19}\text{g cm}^{-3}}\right)^{-\frac{1}{2}}\left(\frac{\mu}{2.3}\right)^{-\frac{3}{2}}
\end{equation}
for $\Theta=1$ (that is, $n=2$). Immediately we notice that the difference between GR and Palatini gravity is given by the solutions of the Lane-Emden equation for different values of the parameter $\alpha$, and hence the Jeans masses for the values from the figure (\ref{fig.1}) are
\begin{equation*}
  M^J_{\alpha=-0.4}/M_\text{GR}=0.96,\;\;\;\;M^J_{\alpha=0.015}/M_\text{GR}=1.003.
\end{equation*}

\subsection{Fragmentation and opacity mass limit}

Currently, there is an agreement that very low-mass stars\footnote{That is, true stars that reached the Main Sequence but with masses below $\sim0.6M_\odot$; to see the effects of modified gravity on such objects, see \cite{aneta2}.}, brown dwarfs, and sub-brown dwarfs (which are sometimes called rogue planets, see e.g. \cite{cab3}) follow the same mechanics of formation, that is, they form via turbulent fragmentation \cite{cab1,cab2,whit}, although there is still room for other processes, depending on the particular case \cite{chab}. Therefore, it turns out that there exist objects below the so-called minimum mass for deuterium burning, which for GR is about $0.0125\pm 0.005M_\odot$ \cite{spiegel} (see \cite{rosyadi} for metric-affine gravity case). That is, those objects form via the fragmentation process but they do not burn any light elements in their cores \cite{kumar1,kumar2}. However, the fragmentation mechanism is restricted by another mass limit, the so-called opacity mass limit ($\sim0.003\pm0.001M_\odot$ for GR \cite{rees}), which is the smallest mass that is bounded gravitationally and that is able to cool via radiation process. It means that it is the smallest mass of a fragment which cannot crumble into smallest pieces caused by gravitational instabilities. Therefore, one may define the opacity mass limit as the minimum mass for a brown dwarf star\footnote{See, however, the nomenclature regarding this topic in \cite{cab3,boss}.}.

For now, we assume that a brown dwarf star\footnote{See, e.g. \cite{burrows1,burrows2}, and \cite{sak1,sak2,gonzalo,maria} for modified gravity models.} is an object between the minimum mass for hydrogen burning and the opacity mass limit; that is, an object which can be massive enough to burn deuterium and eventually lithium, but not massive enough to ignite hydrogen in its core. For GR, brown dwarf stars' mass lies in the mass range $(\sim 0.08-\sim0.003M_\odot)$, depending on the interior structure, the first-order phase transition, opacity and atmosphere model, to mention just a few \cite{aud}.

Using the above result on Jeans mass in Palatini $f(\bar{R})$ gravity for our simplified matter description, we will demonstrate that similarly as it was shown for minimum mass of hydrogen and deuterium burning, the opacity limit can also be affected by a modified gravity model which results as a suchlike degeneracy in the limiting masses.

To begin our analysis, let us consider a fragment's energy rate. It can be shown \cite{stellar} that the characteristic time of the free-fall of the fragment is given by a simple expression
\begin{equation}
   t_\text{free}=(G\rho)^{-\frac{1}{2}}.
\end{equation}
Let us assume that the total energy to be radiated away during collapse is of the order of the gravitational energy $E_g\approx\frac{GM^2}{R}\left(1-\frac{4\alpha}{3\delta}\right)$. Then, in order to keep the fragment always at the same temperature, the rate $A$ of energy to be radiated away is in approximation $A\approx E_g/t_\text{free}$, so:
\begin{equation}
    A=\left(\frac{3}{4\pi}\right)^\frac{1}{2}\frac{G^\frac{3}{2}M^\frac{5}{2}}{R^\frac{5}{2}}\left(1-\frac{4\alpha}{3\delta}\right).
\end{equation}
The maximum luminosity that an object can radiate away is the one of the black-body, given by the Stefan-Boltzmann law (\ref{stefan}). In the case of a planet, such a situation described by this expression would approximately happen when the fragment is already in thermal equilibrium (see the next section for more details). Therefore, the rate of radiation loss for a more realistic case is given by the expression
\begin{equation}
    L=4\pi f\sigma T_\text{eff}^4R^2,
\end{equation}
where $\sigma$ is the Stefan-Boltzmann constant while $f$ is a factor with the value less than one. Introducing it will allow us to take into account that the fragment radiates less than a black-body with the same temperature $T_\text{eff}$.

Considering isothermal collapse, the energy radiated away must be significantly higher than the one of gravitational energy, that is, $L<<A$. When $L\approx A$ we deal with an adiabatic collapse and it will happen for the mass
\begin{equation}
    M^5\approx\frac{64\pi^3}{3}\frac{\sigma^2f^2T_\text{eff}^8R^9}{G^3}\left(1-\frac{4\alpha}{3\delta}\right)^{-2}.
\end{equation}
Using the radius given by (\ref{mass}), eliminating density with the help of (\ref{jeansmass}), and replacing $M$ by $M_J$ in the above relation we obtain the Jeans mass at the end of the fragmentation ($\mu$=1), that is, the opacity limit:
\begin{equation}
    M_J\approx0.003 M_\odot\frac{T_\text{eff}^\frac{1}{4}}{f^\frac{1}{2}}\left(1-\frac{4\alpha}{3\delta}\right)^{-\frac{7}{4}}.
\end{equation}

Therefore, for the given values of the parameter $\alpha$ from the figure \ref{fig.1} the difference between the opacity limits in GR and Palatini gravity again depends on the solutions of the Lane-Emden equation for the given $\alpha$, so 
\begin{equation*}
    M_{\alpha=-0.4}/M_\text{GR}=0.89,\;\;\;\;M_{\alpha=0.015}/M_\text{GR}=1.01.
\end{equation*}
The differences, especially for very small values of the parameter $\alpha$, are not spectacular, one can however also expect disagreement with the models based on GR when a more realistic treatment of the problem is within reach.

\section{Modelling Jovian planets}

The above opacity limit roughly tells us if we deal with a rather (sub-)stellar object such as a brown dwarf star, or a giant planet. As we have demonstrated, modified gravity can also, as in many other cases, introduce an additional degeneracy to limiting masses, which often is a first test determining if one deals with a (brown dwarf) star or a jovian planet. Apart from the mass-relying determination on the nature of the detected object, there exists a designation related to its formation.

It is widely agreed that the planets arise from the gaseous protoplanetary disk surrounding a parent star - from leftovers of the large clouds of gas contracting under its self-gravity. 
There exist two (unnecessarily) mutually exclusive models of the jovian planets' formation: core accretion and disk instability \cite{planet1,planet2,planet3}. The first model requires that such planets can form only in the cool outer region of the protoplanetary disc; after forming a core made of rocky and ice material via two-body collisions and becoming massive enough to trap gas which subsequently collapse onto the planet, the planet starts the cooling process and quasi-equilibrium contraction. On the other hand, the disk instability approach provides a model of formation via gravitational fragmentation of an unstable protoplanetary disk. This process, resulting as a rather massive jovian planet ($\sim 6M_J$) highly depends on cooling time of the contracting fragment and the instability conditions, as well as the way how the energy is transported within the protoplanetary disk. It may also be that both processes are physically viable and happen according to different conditions of a particular star's protoplanetary disk, however it is quite unlikely that the Solar System's giant planets formed via fragmentation.

Nevertheless, in what follows we are interested in the late evolution of the giant gaseous planet which is still undergoing the gravitational contraction. Therefore, we will assume that one of the sources of energy is gravitational contraction apart from radiation received from the parent star \cite{don0}. We will not consider here other internal energy sources such as for instance ohmic heating \cite{oh1,oh2,oh3,oh4} and tidal forces \cite{tid1,tid2,tid3}.

Since the jovian planets' mass is dominated by the contribution from the envelope which surrounds a core of mass $>8 M_\text{Earth}$, in order to understand their physics one needs to understand the physics of gas in different physical conditions. In this work we will deal with a very simple model of a contracting sphere of gas, tracing rather the modified gravity impacts instead of describing realistic interiors and atmospheres of giant planets, for which numerical simulations are usually used.

Summarizing, we will model our jovian planet as a sphere of gas contracting under gravity. The gravitational contraction makes the gaseous planet heat up and radiate this thermal energy. Apart from that, the planet also receives energy from its star - we will also assume here that the planet rotates, allowing it to absorb radiation in equal amounts. We define a planet's atmosphere as a region from which energy is radiated away to space. Using the formalism presented below, we will demonstrate that also the Jupiter's and jovian exoplanets' evolution (which can be also given by the Hertzsprung–Russell diagram) can slightly differ in the framework of modified gravity.

\subsection{Atmosphere quantities for the jovian planets}
As discussed briefly in the section \ref{toolkit}, the planet's luminosity is given by the Stefan-Boltzmann law (\ref{stefan}). Moreover, a given planet can have various energy sources which contribute to the total energy being radiated. Let us assume for now that the only energy source of the planet with the radius $R_\text{p}$ is the energy flux received from the parent star:
\begin{equation}
    L_\text{received}=\left(\frac{R_\text{p}}{2R_\text{sp}}\right)^2L_\text{s},
\end{equation}
where $L_\text{s}$ is the luminosity of the star while $R_\text{sp}$ is the distance between these two objects.
The planet reflects some part of the energy which depends on the planet's albedo $A_\text{p}$; therefore, the energy flux absorbed by the planet is given by the relation
\begin{equation}\label{abs}
    L_\text{abs}=(1-A_\text{p})\left(\frac{R_\text{p}}{2R_\text{sp}}\right)^2L_\text{s}.
\end{equation}
Assuming that the energy absorbed by the planet is uniformly distributed produces the so-called equilibrium temperature $T_\text{eq}$, that is,
\begin{equation}
        (1-A_\text{p})\left(\frac{R_\text{p}}{2R_\text{sp}}\right)^2L_\text{s}=4\pi f\sigma T_\text{eq}^4R_\text{p}^2.
\end{equation}
If the planet is in thermal equilibrium, that is, when the received energy from the parent star balances the one which is radiated away from the planet's surface, gives $T_\text{eq}=T_\text{eff}$. Using that fact (we had assumed before that the only energy source was the one coming from the parent star) and expressing the star's luminosity as $L_\text{s}=4\pi\sigma T_\text{s}^4R_\text{s}^2$, we may write down the equilibrium temperature as (for more  details see \cite{don})
\begin{equation}\label{teq}
  T_\text{eq}=  (1-A_\text{p})^\frac{1}{4}\left(\frac{R_\text{s}}{2R_\text{sp}}\right)\frac{1}{2}T_\text{s}.
\end{equation}
Let us notice that the equilibrium temperature does not depend on the planet's radius when the only energy source is that of the parent star. However, it is not true when there exist other (internal) energy sources, such as already mentioned gravitational contraction, Ohmic heating, or tidal forces. Because of that fact the planet radiates away more energy than it receives from the parent star, and consequently its effective temperature is higher than the equilibrium one. 

In order to find the relation between effective and equilibrium temperatures, one uses the standard equation for radiative transfer in grey atmosphere \cite{saumon,guillot, seager} together with Eddington's approximation\footnote{See e.g. \cite{hansen}.}. Therefore, it can be shown that \cite{don} 
\begin{equation}\label{temp}
 4T^4=3\tau(T^4_\text{eff}-T^4_\text{eq})+2(T^4_\text{eff}+T^4_\text{eq}),
\end{equation}
where $T$ is the stratification temperature in the atmosphere while $\tau$ is the optical depth given by (\ref{eq:od}). The optical depth is zero at the surface of the planet which is used as a boundary condition to get (\ref{temp}). For simplicity, we will use the following abbreviations in the further parts of the paper:
\begin{equation*}
    T_-:=T^4_\text{eff}-T^4_\text{eq},\;\;\;T_+:=T^4_\text{eff}+T^4_\text{eq}
\end{equation*}
such that the equation (\ref{temp}) is now
\begin{equation}\label{temp2}
 4T^4=3\tau T_-+2T_+.
\end{equation}

The atmosphere is in hydrostatic equilibrium with the gravitational pressure - hence we will use that fact to find the pressure in the atmosphere.
As mentioned, the optical depth definition is a useful tool to integrate the hydrostatic equilibrium equation (\ref{pres}) and get a relation for the pressure at the atmosphere. Therefore, using (\ref{pres}) and (\ref{surf}) one can write
\begin{equation}\label{equil}
 \frac{dp}{dr}=-\kappa\rho\frac{dp}{d\tau}=-g\rho\left(1-\frac{4\alpha}{3\delta}\right).
\end{equation}
Since we are dealing with low temperatures, thus the opacity can be written in a simple power-low, that is,
\begin{equation}\label{opacity}
 \kappa=\kappa_0 p^u T^{4w},
\end{equation}
where the value of $\kappa_0$ depends on various opacity regimes by type of matter the atmosphere consists of \cite{kley,armitage}. The powers $u$ and $w$ are values related to the energy transport in the envelope and they will be kept general for now.
Using the formula (\ref{opacity}), the equation (\ref{equil}) is now
\begin{equation}\label{prestemp}
 p^u \frac{dp}{d\tau}=\frac{g}{\kappa_0 T^{4w}}\left(1-\frac{4\alpha}{3\delta}\right).
\end{equation}
Substituting the expression (\ref{temp}) to the above
\begin{equation}
 \int^p_0 p^u dp=\frac{4^w g}{\kappa_0}\left(1-\frac{4\alpha}{3\delta}\right)\int^\tau_0\frac{d\tau}{(3\tau T_-+2T_+)^w}
\end{equation}
we may integrate it for $w\neq1$ and $w=1$, respectively, to find the atmospheric pressure of the form:
\begin{align}\label{presat}
 p^{u+1}=&\frac{4^wg}{3\kappa_0}\frac{u+1}{1-w}\left(1-\frac{4\alpha}{3\delta}\right)\nonumber\\
 \times&T_-^{-1}\Big((3\tau T_-+2T_+)^{1-w}-(2T_+)^{1-w}\Big),\\
 p^{u+1}=&\frac{4g}{3\kappa_0}(u+1)\left(1-\frac{4\alpha}{3\delta}\right)T_-^{-1}\text{ln}[3\tau T_-+2T_+],
\end{align}
where the boundary condition $p=0$ at $\tau=0$ has been used.

\subsection{Boundary between radiative atmosphere and convective interior}
Interiors of gaseous giant planets, as well as of brown dwarfs \cite{burrows1,burrows2}, are fully convective, that is, energy is transported by convective processes. Therefore, there exists a region between the interior and the atmosphere where convection is replaced by radiative transport. The condition for this change is given by the Schwarzschild criterion, described briefly in the section \ref{toolkit} after the equation (\ref{hyd}). The convective interior is well described by the polytropic equation of state (\ref{pol}) with $n=3/2$ and hence for the fully ionised gas stratification $d\ln{T}/d\ln{p}=\nabla_\text{ad}$ is adiabatic and equaled to $2/5$ \cite{stellar}. 
Using the equation (\ref{prestemp}) together with $ \frac{dp}{d\tau}=\frac{dp}{dT}\frac{dT}{d\tau}$, and applying the Schwarzschild condition to it one has
\begin{equation}
    \frac{15}{36}p^{u+1}T^{-4}T_-=\frac{g}{\kappa_0 T^{4w}}\left(1-\frac{4\alpha}{3\delta}\right).
\end{equation}
Substituting the temperature of the atmosphere (\ref{temp2}) and atmospheric pressures (\ref{presat}) we find that the critical depth is
\begin{align}
 \tau_c&=\frac{2}{3}\frac{T_+}{T_-}\left(\Big(1+\frac{8}{5}\Big(\frac{w-1}{u+1}\Big)\Big)^\frac{1}{w-1}-1\right),\;\;w\neq1\\
 \tau_c&=\frac{2}{3}\frac{T_+}{T_-}(e^\frac{16}{15}-1),\;\;w=1.
\end{align}

This is the optical depth at which the radiative transport is replaced with the convective one. Let us notice that those expressions do not depend on Palatini gravity; they have the same form as in \cite{don}. Then, in order to find pressure and temperature $T$ at the boundary between radiative atmosphere and convective interior one needs to substitute those relations to (\ref{temp2}) and (\ref{presat}), respectively:
\begin{align}
 p^{u+1}_\text{conv}=&\frac{8g}{15\kappa_0}\frac{4^w\left(1-\frac{4\alpha}{3\delta}\right)}{T_-(2T_+)^{w-1}}\left(\frac{5(u+1)}{5u+8w-3}\right),\label{p1}\\
 T^4_\text{conv}=&\frac{T_+}{2} \left(\frac{5u+8w-3}{5(u+1)}\right)^{w-1}
\end{align}
for $w\neq1$ while for $w=1$ those equations reduce to
\begin{align}
 p^{u+1}_\text{conv}=&\frac{32g}{15\kappa_0}\frac{\left(1-\frac{4\alpha}{3\delta}\right)}{T_-},\\
 T^4_\text{conv}=&\frac{1}{2}T_+e^\frac{16}{15}.
\end{align}

\subsection{Convective interior of the jovian planets}
Jupiter-like planets have still a lot to uncover, but what one can say for sure is that those majestic giants possess a complex internal structure. Starting with theoretical works,  \cite{jeff,jup,jup1,jup2,jup3,jup4,jup5} and the revelations provided by Juno mission on Jupiter's interior \cite{juno,juno2,juno3,stev_juno}, the up-to-date model of the planet consists of at least three layers with not sharp boundaries, in contrary to the terrestrial planets: a (possibly diffusive) core made of heavier elements, a mantle, in which the dominant element is metallic hydrogen with some abundances of helium and heavier elements; and a molecular hydrogen envelope with helium rain and silicate droplets \cite{helled}. Another issue is related to the behaviour of hydrogen and hydrogen-helium mixture in the pressure near to megabar at a few thousands degrees since it is not well tested yet \cite{stev_juno}, and hence some changes in the common accepted equations of state can be also necessary \cite{mix1,mix2}.

\begin{figure}[t]
\centering
\includegraphics[scale=.8]{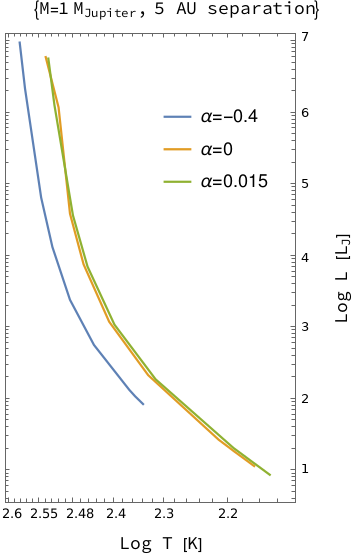} 
\caption{[color online] The Hertzsprung–Russell diagram for a jovian planet at the $5$AU distance from its parent star for a few values of the parameter $\alpha$. Each curve represents an evolution of a Jupiter-mass planet, starting from the radius $R=\sim10^{10}$ to the radius $R=\sim10^8$. }
\label{fig.2}
\end{figure}

Remembering the difficulties regarding the accurate description of the Jupiter and jovian planets, in what follows, we will model their interior pressure by the simplified combination \cite{don0}
\begin{equation}\label{prescomb}
    p=p_1+p_2,
\end{equation}
where $p_1$ is pressure arising from electron degeneracy, given by the polytropic equation of state (\ref{pol}) with $n=3/2$,
while $p_2$ is pressure of ideal gas
\begin{equation}
    p_2=\frac{k_B\rho T}{\mu},
\end{equation}
where $\mu$ is the mean molecular weight. It turns out, however, that such a combination can be written as a new polytropic equation of state with $n=3/2$ \cite{stevenson}, say, $p=A\rho^\frac{5}{3}$, such that $\rho=(p/A)^\frac{3}{5}$, where $A=p_c/\rho_c^\frac{5}{3}$. Substituting this to the pressure combination (\ref{prescomb}) and using the relations from the Lane-Emden formalism given in the section \ref{toolkit} to write
\begin{equation}
 A=\gamma^{-1}GM_p^\frac{1}{3}R_p,
\end{equation}
we may express the interior pressure (\ref{prescomb}) as
\begin{equation}\label{p2}
 p_\text{conv}=\frac{GM_p^{1/3}R_p}{\gamma}\left(\frac{kT_\text{conv}}{\mu\left(G\gamma^{-1}M_p^{1/3}R_p-K\right)}\right)^\frac{5}{2}.
\end{equation}
The obtained pressure (\ref{p2}) must be equal to the boundary pressure (\ref{p1}) providing us an equation which relates the effective temperature $T_\text{eff}$ with the radius of the planet $R_p$:
\begin{align}\label{cond}
 &T_+^{\frac{5}{8}u+w-\frac{3}{8}}T_-=CG^{-u} M_p^{\frac{1}{3}(2-u)}R_p^{-(u+3)}\mu^{\frac{5}{2}(u+1)}k_B^{-\frac{5}{2}(u+1)}\nonumber\\
 &\times\gamma^{u+1}(G\gamma^{-1}M_p^\frac{1}{3}R_p-K)^{\frac{5}{2}(u+1)}\left(1-\frac{4\alpha}{3\delta}\right)
\end{align}
where $C$ is a numerical constant which depends on the opacity constants $u$ and $w$:
\begin{align}
 C_{w\neq1}=&\frac{16}{15\kappa_0}2^{\frac{5}{8}(1+u)+w}\left(\frac{5u+8w-3}{5(u+1)}\right)^{1+\frac{5}{8}(1+u)(w-1)},\\
 C_{w=1}=&\frac{32}{15\kappa_0}2^{\frac{5}{8}(u+1)}e^{-\frac{2}{3}(u+1)}.
\end{align}
The equation (\ref{cond}) is valid for all values $w>1$. When the planet's contraction is over, the only source of energy heating the planet is the parent star; it means that the effective temperature $T_\text{eff}$ is equaled to the equilibrium one $T_\text{eq}$, so $T_-=0$. That gives the final radius $R_F$ from the relation (\ref{cond})
\begin{equation}
 R_F=\frac{K\gamma}{GM_p^\frac{1}{3}},
\end{equation}
which slightly differs in Palatini gravity with respect to General Relativity by the value of $\gamma$. The equation (\ref{cond}) can be solved numerically in the radius range $\sim(10^{10};\,10^8)$ for the given values of parameter $\alpha$ and equilibrium temperature (\ref{teq}) in order to get the effective temperature $T_\text{eff}$ for each $R_p$ during the contraction. The constants related to the opacity model are taken from \cite{don} while the luminosity is obtained from Stefan-Boltzmann law (\ref{stefan}). As a reference planet's mass we took Jupiter's, therefore the distance between the parent star, that is, the Sun and the planet, is $\sim5$AU. The results are presented in the figure \ref{fig.2} for three different values of the parameter $\alpha$; the GR/Newtonian model is given by $\alpha=0$.

\subsection{Jovian planets' evolution}

Assuming that the contraction of the planet is a quasi-equilibrium process, we may write down the luminosity of the planet which is a sum of the total energy absorbed by the planet, $L_\text{abs}$, and the internal energy whose source is the gravitational energy. Thus, for a polytrope with $n=3/2$ \cite{maria}, we have
\begin{equation}\label{cooling}
 L_p=L_\text{abs}-\frac{3}{7}\frac{GM_p^2}{R_p^2}\frac{dR_p}{dt}.
\end{equation}
Using the previous formulas (\ref{stefan}) and (\ref{abs}) we may write the evolution equation (\ref{cooling}) as
\begin{equation}
 \pi a c R^2_pT_-=-\frac{3}{7}\frac{GM_p^2}{R_p^2}\frac{dR_p}{dt},
\end{equation}
which can be integrated from an initial radius $R_0$ to the final one $R_F$, providing the timescale for contraction\\
\begin{equation}
 t=-\frac{3}{7}\frac{GM_p^2}{\pi ac}\int^{R_p}_{R_0}\frac{dR_p}{R_p^4T_-}.
\end{equation}
Using the equation (\ref{cond}), we may get rid of $T_-$ and write the cooling equation as a function of the opacity parameters
\begin{align}\label{age}
 t=&-\frac{3}{7}\frac{GM_p^\frac{4}{3}k_B^{\frac{5}{2}(u+1)}\kappa_0}{\pi ac\gamma\mu^{\frac{5}{2}(u+1)}K^{\frac{3}{2}u+\frac{5}{2}}C}\left(1-\frac{4\alpha}{3\delta}\right)^{-1}\nonumber\\
 &\times\int^{x_p}_{x_0}\frac{(T_\text{eff}^4+T^4_\text{eq})^{\frac{5}{8}u+w-\frac{3}{8}}dx}{x^{1-u}(x-1)^{\frac{5}{2}(u+1)}}.
\end{align}

Let us observe that it takes an infinite time to reach the thermal equilibrium, independently of the model of gravity. However, to reach a particular stage of evolution, a jovian planet can be younger/older than predicted in General Relativity, as signalized by the $\alpha$-depending term in (\ref{age}), as well as by different values of the effective temperature given by (\ref{cond}). A few values of the effective temperature, luminosity, and the age ratios are given in the tables (\ref{tab1}) and (\ref{tab2}) for $\alpha=-0.4$ and $\alpha=0.015$, respectively.

\begin{table}
\caption{The effective temperature, luminosity (in Jupiter's luminosity), and age ratio with respect to the GR values at the given point of the jovian planet's evolution for the given radius $R$ for $\alpha=-0.4$.}
\centering
\begin{tabular}{llll }
\hline\noalign{\smallskip}
R ($10^9$m) & $T_\text{eff}$ (K) & L/L$_\text{J}$  & $t_\alpha/t_\text{GR}$ \\
\noalign{\smallskip}\hline\noalign{\smallskip}
10 & 381 & 9$\times10^6$ & 1.5\\
5 & 372 & 2$\times10^6$ & 1.7 \\
1 & 346 & 0.6$\times10^5$ & 1.6 \\
0.5 & 329 & 13$\times10^3$ & 1.7 \\
0.25 & 304 & 2$\times10^3$ & 1.9 \\
0.15 & 273 & 5.6$\times10^2$ & 2.4 \\
0.1 & 233 & 1.3$\times10^2$ & 3.8 \\
0.095& 227 & 1.1$\times10^2$ & 4.3 \\
0.09 & 219 & 83 & 4.7 \\ 
\noalign{\smallskip}\hline
\end{tabular}\label{tab1}
\end{table}

\begin{table}
\caption{The effective temperature, luminosity (in Jupiter's luminosity), and age ratio with respect to the GR values at the given point of the jovian planet's evolution for the given radius $R$ for $\alpha=0.015$.}
\centering
\begin{tabular}{llll }
\hline\noalign{\smallskip}
R ($10^9$m) & $T_\text{eff}$ (K) & L/L$_\text{J}$  & $t_\alpha/t_\text{GR}$ \\
\noalign{\smallskip}\hline\noalign{\smallskip}
10 & 325 & 5.6$\times10^6$ & 0.96\\
5 & 326 & 1.3$\times10^6$ & 1.1 \\
1 & 300 & 0.4$\times10^5$ & 0.95 \\
0.5 & 281 & 7$\times10^3$ & 0.93 \\
0.25 & 249 & $10^3$ & 0.92 \\
0.15 & 207 & 2$\times10^2$ & 0.89 \\
0.1 & 146 & 0.2$\times10^2$ & 0.83 \\
0.095& 135 & 0.1$\times10^2$ & 0.83 \\
0.09 & 124 & 8.5 & 0.85 \\ 
\noalign{\smallskip}\hline
\end{tabular}\label{tab2}
\end{table}

\section{Conclusions}
In the presented paper we have studied the opacity mass limit which is, roughly speaking, a boundary mass between brown dwarf stars and giant gaseous (exo-)planets, and the evolution of the last ones in the framework of the quadratic, $f(\bar{R})=\bar{R}+\beta\bar{R}^2$, Palatini gravity. For the purposes of this work, we have derived a simplified version of the Jeans mass; this critical mass is slightly smaller (bigger) for positive (negative) parameter $\beta$, respectively\footnote{Let us recall that $\alpha=\kappa c^2 \beta\rho_c$, where $\kappa$ is defined due to the negative sign convention.}. This means that modified gravity will also have an impact on the process of fragmentation and its limit, that is, the opacity mass limit. As expected, the minimal mass for a brown dwarf is altered in the similar manner as Jeans mass, and therefore Palatini $f(\bar{R})$ gravity as well as any other theory of gravity which even slightly modifies the Newtonian limit, introduces the additional uncertainty to the substellar objects' classifications which rely on limiting masses. 

Regarding the evolution of the jovian planets we have also expected small differences with respect to the models which are based on Newtonian gravity. Let us firstly notice that we have altered only the planetary structure equations, without taking into account the effects of modified gravity on the parent star (as a reference distance between the jovian planet and parent star, as well as the star's and planet's properties we have taken the Sun-Jupiter system). It turns out that Palatini gravity introduces an extra term to the constraint relating the radius and effective temperature of the planet (\ref{cond}), as well as to the evolutionary equation (\ref{age}), from which we can obtain the age of the planet at any point of the Hertzsprung–Russell diagram (\ref{fig.2}). Some of those values, that is, the effective temperature, luminosity (obtained from the equation (\ref{stefan}) for the given radius), and the age's ratio with respect to the Newtonian model for the Jupiter-like planet are given in the tables \ref{tab1} and \ref{tab2}. In the case of the temperatures and luminosities there are not very big differences with respect to the Newtonian models (by saying this we take into account the possible observational uncertainties, together with the assumptions and simplifications of our model), although we observe a shift of the evolutionary curve positions on the H-R diagram (\ref{fig.2}) with respect to the different values of the parameter $\alpha$. However, there are significant changes in the ages of the planet when it contracts, especially in the late stage of the evolution. This means that the giant planets of our Solar System, Jupiter and Saturn, can be much older (positive $\beta$ parameter) or younger (negative $\beta$) than we have believed so far, obtaining their ages by using Newtonian gravity. This result may change our current knowledge on the Solar System formation since many processes, such as, for instance, considered here fragmentation, but also the early stellar evolution \cite{aneta2}, cooling of substellar objects \cite{maria}, and planets' profiles \cite{olek}, differ in the framework of modified gravity. 

Let us comment that although we have presented a very simplified analysis of some of the processes which happen in the Solar System, models and simulations which take into account physics which we have not considered in the paper\footnote{Such as, for instance, rotation \cite{rot}, magnetic fields \cite{magn}, complex inner structure \cite{core}, gas description \cite{manuel,manuel2,kremer3}, ...}, are based on Newtonian equations. Because of that reason we also expect that in a more realistic approach we will deal with an altered description when other model of gravity is applied, and our understanding of processes occurring in our nearest neighbourhood may also change. The research in this direction is on high demand, especially in the light of many current and future missions, whose aim is to explore our and other planetary systems, and to provide more accurate data regarding the substellar objects \cite{vision,voyage,webb,nancy,tess,spitzer,nn}.

\vspace{5mm}
\noindent \textbf{Acknowledgement.} 
This work was supported by the EU through the European Regional Development Fund CoE program TK133 ``The Dark Side of the Universe." The author would like to thank the members of the National Institute of Chemical Physics and Biophysics group in Tallinn, Estonia, for their hospitality during the initial part of this work.

\end{document}